\documentclass[aps,preprint,showpacs,superscriptaddress, groupedaddress]{revtex4}  % for double-spaced preprint
\usepackage{graphicx}  % needed for figures
\usepackage{dcolumn}   % needed for some tables
\usepackage{bm}        % for math
\usepackage{amssymb}   % for math
\usepackage{lineno}

\begin{document}

% the following line is for submission, including submission to the arXiv!!
%\hspace{5.2in} \mbox{Fermilab-Pub-04/xxx-E}

\title{{Dynamical System's approach to relativistic nonlinear wave-particle interaction in weakly collisional plasmas.}}
%\author{A. Osmane\footnote{a.osmane@unb.ca} and A.M. Hamza}
\author{A. Osmane and A. M. Hamza}
\affiliation{Department of Physics, University of New Brunswick, Fredericton, NB, E3B5A3, CANADA}
\email{a.osmane@unb.ca}
\date{\today}

\begin{abstract}
In this report, we present a dynamical systems' approach to study the exact nonlinear wave-particle interaction in relativistic regime. We give a particular attention to the effect of wave obliquity on the dynamics of the orbits by studying the specific cases of parallel ($\theta=0$) and perpendicular ($\theta=-\pi/2$) propagations in comparison to the general case of oblique propagation $\theta=]-\pi/2, 0[$. We found that the fixed points of the system correspond to Landau resonance, and that the dynamics can evolve from trapping to surfatron acceleration for propagation angles obeying a Hopf bifurcations condition. Cyclotron-resonant particles are also studied by the construction of a pseudo-potential structure in the Lorentz factor $\gamma$. We derived a condition for which Arnold diffusion results in relativistic stochastic acceleration. Hence, two general conclusions are drawn : 1) The propagation angle $\theta$ can significantly alter the dynamics of the orbits at both Landau and cyclotron-resonances. 2) Considering the short-time scales upon which the particles are accelerated, these two mechanisms for Landau and cyclotron resonant orbits could become potential candidates for problems of particle energization in weakly collisional space and cosmic plasmas.  
\end{abstract}

%\pacs{94.05.-a, 94.05.Pt, 94.30.Xy}
\maketitle

\section{Introduction} The wave-particle interaction has long been considered a dominant energy-momentum exchange mechanism in space and astrophysical plasmas. Beyond the dense and internal boundaries of stars and planetary magnetospheres, space and astrophysical plasmas are predominantly collisionless, and populated by distributions functions inconsistent with collisional equilibrium conditions. These plasmas are also believed to be turbulent systems described by a conventional energy cascade from large scales to small scales where dissipation takes place. While fluid theories provide satisfactory descriptions of macroscopic quantities at large scales, they are not equipped to explain the plasma physics at smaller kinetic scales, and one needs to include nonlinear kinetic processes (wave-particle interactions and wave-wave interactions) for a correct description of these turbulent and collisionless plasmas \cite{Kulsrud}.\\
Studies of wave-particle interactions in space and astrophysical turbulent plasmas have commonly fallen under the scope of quasi-linear theory \cite{Kennel66, Roux70}. Quasi-linear theory departs from linear theory in conserving energy and momentum through the account of the wave-particle interaction, resulting in a diffusion process for an ensemble averaged distribution function solution to a Fokker-Planck equation. However, quasi-linear theory is constrained by a number of severe limitations making it inapplicable for plasmas containing large-amplitude quasi-monochromatic electromagnetic and/or electrostatic waves. Indeed, the first assumption for quasi-linear theory consists in constraining particle orbits to their unperturbed components. A second assumption consists in assuming a wave spectrum sufficiently dense so that interference between modes is destroyed by phase-mixing. Hence, quasi-linear theory is valid only when the bandwidth is broad enough to enable resonances to be maintained as particles are scattered, and non-linear trapping effects by individual waves are to weak to be taken into consideration. \\%Paragraph on the Radiation belts
Due to the inherent difficulties of strong turbulence theories, test-particle methods have become a favored tool for the study of wave-particle interaction beyond the constraints of quasi-linear theory. Numerous methods have been developed in the context of the radiation belts alone, ranging from guiding-center approximation \cite{Omura82}, to resonance-averaged Hamiltonian \cite{Albert02}, gyroresonance averaged equations \cite{Bortnik08} and computer simulations taking into account approximate and exact dipolar fields [see review \cite{Sh08} and associated references]. However, the general case of oblique has often been avoided in favor of the more tractable case of parallel of propagation \cite{Tao10}. A strong case can be made for the neglect of oblique propagation and nonlinear effects for small amplitude waves since the appearance of a small parallel electric field can not trap orbits \cite{Albert10}. However, if the electric field component of the wave becomes sufficiently large, such that nonlinear effects can be triggered, then a rich class of orbits can result, and the parallel approximation becomes invalid.\\    
Recent observations of the weakly collisional plasma of the radiation belts are suggesting that obliquely propagating waves with Poynting flux two orders of magnitude larger than previously observed whistlers waves are commonly generated in the radiation belts and appear correlated with relativistic electron microbursts \cite{Wilson11}. These large amplitude waves can propagate at large propagation angles ($\theta \leq 70^o$), and possess amplitudes capable of energizing electrons on time scales of the order of the milliseconds \cite{Yoon11}. If these large-amplitudes waves are shown to be a common observational signature in the radiation belts, the conventional models used to describe the wave-particle interaction using a quasi-linear formalism will have to be revisited. It is not only inaccurate to assume that a particle will execute a random walk in pitch angle during the course of one bounce period, but as demonstrated below, a particle can be irreversibly accelerated to relativistic energies in less than one bounce period.\\
In this report, we investigate the exact nonlinear wave-particle interaction in the relativistic regime. The inclusion of relativistic effects is a \textit{sine qua non} condition for any attempt at solving the outstanding problems which have emerged in radiation belts dynamics as well as galactic cosmic ray.  Our goal is to provide a general framework for the wave-particle interaction by using a dynamical systems' approach. Such approach, although lacking the level of self-consistency found in numerical simulations, can facilitate the understanding of complex systems such as cosmic and space plasmas and therefore provide for an intuitive as well as quantitative leap between theoretical models and simulations.\\
The report is written as follows. In section 2 we derive the dynamical system as well as its fixed points and invariants. In section 3 we treat the special cases of parallel and purely perpendicular propagation. In section 4 we study the general case of oblique propagation for the cyclotron-resonance case as well as the Landau resonance. Section 5 contains a discussion of the general framework for the understanding of the wave-particle interaction and the effect of oblique propagation in collisionless plasmas such as the radiation belts. In section 6 we conclude and discuss studies currently underway to address limitations of the dynamical system approach.

\section{Dynamical System.}
\subsection{Equation of motion for the general case.} 
Our study begins with the equation of motion of a particle in an electromagnetic field, as described by the Lorentz equation. The equations of motion can be written as
\begin{equation}
\frac{d\textbf{p}}{dt}= e\bigg{[}\textbf{E}(\textbf{x},t)+\frac{\textbf{p}}{m\gamma c} \times \textbf{B}(\textbf{x},t)\bigg{]},
\end{equation}
for a particle of charge $e$, momentum $\textbf{p}=m\gamma\textbf{v}$ and rest mass $m$. The Lorentz factor, $\gamma$, is defined in terms of the relativistic momentum as follows : 
\begin{equation}
\label{eq:constraint}
\gamma=\sqrt{1+\frac{p^2}{m^2c^2}}.
\end{equation}
The electromagnetic field configuration consists of a background magnetic field $B_0$ to which is superposed an electromagnetic wave given by $(\delta \textbf{E}, \delta \textbf{B})$ : 
\begin{equation}
\textbf{E}(\textbf{x},t)=\delta \textbf{E}(\textbf{x},t)
\end{equation}
\begin{equation}
\textbf{B}(\textbf{x}.t)=\textbf{B}_0+\delta \textbf{B}(\textbf{x},t)
\end{equation}
The electromagnetic wave vector $\mathbf{k}$ is chosen to point in the $\hat{z}$ direction, obliquely to the background magnetic field lying in the y-z plane : 
\begin{equation}
\textbf{k} \cdot \textbf{B}_0=kB_0\cos(\theta)
\end{equation}
\begin{equation}
\left\{ 
\begin{array}{l l} 
\delta\textbf{E} =  \delta E_x \hat{\textbf{x}} + \delta E_y \hat{\textbf{y}}\\ 
\delta\textbf{B} =  \delta B_x \hat{\textbf{x}} + \delta B_y \hat{\textbf{y}}\\   
\end{array} \right.
\end{equation}
with the wave magnetic field components written as 
\begin{equation}
\left\{ 
\begin{array}{l l} 
\delta B_x= \delta B \sin(kz-\omega t)\\ 
\delta B_y=\delta B \cos(kz-\omega t)\\   
\end{array} \right.
\end{equation}
and Faraday's law, expressed in terms of the Fourier components, providing for the components of the electric field\begin{equation}
c\textbf{k} \times \delta \textbf{E} (\textbf{k},\omega) =\omega \delta \textbf{B}(\textbf{k},\omega)
\end{equation}
We can therefore express the dynamical system in terms of the phase velocity $v_\phi=\omega/k$,  and the variables $p_\Phi=m\gamma v_\Phi$ ; $\Omega_1=e\delta B/mc\gamma$ ; $\Omega_0=e B_0/mc\gamma$, resulting in the following coupled ordinary differential equations : 
\begin{equation}
\left\{ 
\begin{array}{l l} 
\dot{p}_x=p_y\Omega_0\cos(\theta)+(p_\Phi-p_z)\Omega_1\cos(kz-\omega t)  +p_z\Omega_0\sin(\theta)\\
\dot{p}_y=-p_x\Omega_0\cos(\theta)+(p_z-p_\Phi)\Omega_1\sin(kz-\omega t)\\
\dot{p}_z=-p_x\Omega_0 \sin(\theta)+p_x\Omega_1\cos(kz-\omega t) -p_y\Omega_1\sin(kz-\omega t)\\
\dot{z}=p_z v_\Phi/p_\Phi
\end{array} 
\right.
\end{equation} 
In the classical case, the dynamical system is composed of four equations, the three components of the velocity plus the position coordinate along $k$. However, in the relativistic case the expression for the Lorentz factor must be obeyed and constitutes a constraint on the particle's trajectory.  We can keep track of this constraint by adding an equation in the expression of a dynamical gyrofrequency : 
\begin{eqnarray}
\label{eq:omega_dot}
\dot{\Omega}_0 & = & \frac{d}{dt}\bigg{(}\frac{eB_0}{mc\gamma}\bigg{)} \nonumber \\
                              &=&-\Omega_0 \frac{pc^2}{m^2c^4+p^2c^2}\dot{p}\nonumber\\
                            &=&-\frac{\Omega_0\Omega_1p_\Phi}{m^2\gamma^2c^2}\bigg{(}p_x\cos(kz-\omega t)-p_y\sin(kz-\omega t)\bigg{)}
\end{eqnarray}
If we define the constant $\delta_1=\Omega_1/\Omega_0$, it is straightforward to see that $\dot{\Omega}_1=\delta_1 \dot{\Omega}_0$, and similarly, since $p_\Phi=p_\Phi(\gamma)$, the time evolution of this quantity can be written as :
\begin{equation}
\label{eq:p_phi}
\dot{p}_\Phi=-mv_\Phi\gamma\frac{\dot{\Omega}_0}{\Omega_0}.
\end{equation} 
In order to simplify the dynamical system, we can eliminate the explicit time dependence of the equations by making the following mathematical transformation :
\begin{equation}
p_x'=p_x, \hspace{2mm}p_y'=p_y, \hspace{2mm} p_z'=\gamma_w(p_z-p_\phi), \hspace{2mm} z'=\gamma_w(z-v_\phi t)
\end{equation}
for the Lorentz factor :
\begin{equation}
\gamma_w=\frac{1}{\sqrt{1-\frac{v_\Phi^2}{c^2}}}.
\end{equation}
We can therefore write the equations of motion as follow :
\begin{equation}
\left\{ 
\begin{array}{l l} 
\dot{p}_x'=\Omega_0 p_y'\cos(\theta)-\Omega_1p_z' \cos(kz'/\gamma_w)/\gamma_w  +\Omega_0 (p_z'/\gamma_w+p_\phi) \sin(\theta)\\
\dot{p}_y'=-\Omega_0p_x'\cos(\theta)+\Omega_1p_z' \sin(kz'/\gamma_w)/\gamma_w\\
\dot{p}_z'/\gamma_w=-\Omega_0p_x'\sin(\theta)+\Omega_1p_x'\cos(kz'/\gamma_w) -\Omega_1p_y'\sin(kz'/\gamma_w)-\dot{p}_\Phi\\
\dot{z}'=p_z'v_\Phi/p_\Phi\\
\end{array} 
\right.
\end{equation} 
If we absorb the Lorentz factor $\gamma_w$ into $p_z'$ and $k$, that is we write $p_z' \to p_z'/\gamma_w$ and $k \to k/\gamma_w$, and write $\dot{p}_\Phi$ in terms of $(p_x',p_y',p_z',z',\Omega_0)$, we can express the dynamical system as : 
\begin{equation}
\label{eq:ds_in_ps}
\left\{ 
\begin{array}{l l} 
\dot{p}_x'=\Omega_0 p_y'\cos(\theta)-\Omega_1p_z' \cos(kz')  +\Omega_0 (p_z'+p_\phi) \sin(\theta)\\
\dot{p}_y'=-\Omega_0p_x'\cos(\theta)+\Omega_1p_z' \sin(kz')\\
\dot{p}_z'=-\Omega_0p_x'\sin(\theta)+\Omega_1(\frac{n^2-1}{n^2})(p_x'\cos(kz') -p_y'\sin(kz'))\\
\dot{z}'=p_z'v_\Phi/p_\Phi\\
\end{array} 
\right.
\end{equation} 
where the refractive index is represented as $n^2=c^2/v_\Phi^2$. The magnitude of the momentum is now written as $p'=\sqrt{p_x^{'2}+p_y^{'2}+(p_z^{'}/\gamma_w)^2}$, hence the Lorentz contraction factor also transforms from $\gamma(p) \to \gamma(p')$. The dynamical system for the classical case can be recovered by setting $\gamma = 1$ and $1/n^2\rightarrow 0$. Indeed,  the equations are equivalent to the classical case under the following transformations : $p \to u$ and $\Omega \to \Omega/\gamma$ . Hence the main difference lies in the time dependence of the Larmor frequencies and the extra term that goes as $1/n^2$ in the $\dot{p}_z'$ equation.

\subsection{Representation in terms of ($P$, $\alpha$, $\Phi$, $z'$).}
It is convenient to express the relativistic momentum in spherical coordinates, that is in terms of a magnitude $p'$ and phase angles $(\alpha, \Phi)$. This can be achieved by introducing the following scalar and vector variables :
\begin{equation}
\left\{ 
\begin{array}{l l} 
p_{\parallel} = \textbf{p}\cdot \hat{\textbf{b}}_0\\
\textbf{p}_\perp=\hat{\textbf{p}}_0 \times (\textbf{p} \times \hat{\textbf{p}}_0) = \textbf{p}-p_{\parallel}\hat{\textbf{b}}_0\\
\end{array} 
\right.
\end{equation} 
where $\hat{\textbf{b}}_0 = \textbf{B}_0/B_0$. Using these definitions, we can rewrite the momentum $\textbf{p}'=(p_x',p_y',p_z')$ in terms of the pitch angle $\alpha$ and the dynamical gyrophase  $\Phi$, both defined as

\begin{equation}
\left\{ 
\begin{array}{l l} 
\label{eq:tan_alpha}
\tan (\alpha) = {\displaystyle\frac{p'_\perp}{p'_{\parallel}}}\\
\tan (\Phi) ={\displaystyle\frac{p'_{\perp 1}}{p'_{\perp 2}}=\frac{p_x'}{p_y'\cos(\theta)+p_z'\sin(\theta)}}
\end{array} 
\right.
\end{equation}

Hence, all three momentum components in the wave frame are written as : 
\begin{equation}
\label{eq:spherical}
\left\{ 
\begin{array}{l l} 
p_x'=p'\sin(\alpha)\cos(\Phi)\\
p_y'=p'\sin(\alpha)\sin(\Phi)\cos(\theta)-p'\cos(\alpha)\sin(\theta)\\
p_z'=p'\sin(\alpha)\sin(\Phi)\sin(\theta)+p'\cos(\alpha)\cos(\theta).\\
\end{array} 
\right.
\end{equation}
Using the definitions (\ref{eq:tan_alpha}) and the representation of the momentum in (\ref{eq:spherical}), we can proceed to write the dynamical system (\ref{eq:ds_in_ps}) in terms of the normalized variables :  
\begin{equation}
\label{eq:norm}
P=\frac{kp'}{m\omega},\hspace{1mm}\delta_1=\frac{\Omega_1}{\Omega_0},\hspace{1mm}\delta_2=\frac{m\omega c}{eB_0}, \hspace{1mm}\delta_3=\frac{1}{\delta_2\gamma}, \hspace{1mm}n^2=\frac{c^2}{v_\Phi^2},\hspace{1mm} Z=kz', \tau=\omega t
\end{equation}
and the function $F(\alpha, \Phi, Z)$, as follow : 
\begin{equation}
\label{eq:DS}
\left\{ 
\begin{array}{l l} 
{\displaystyle\frac{dP}{d\tau}}=\sin(\alpha)\cos(\Phi)\sin(\theta)/\delta_2 \\
                   \hspace{8mm}- {\displaystyle\frac{\delta_1\delta_3 P}{n^2}}\bigg{(}\sin(\alpha)\sin(\Phi)\sin(\theta)+\cos(\alpha)\cos(\theta)\bigg{)}F(\alpha, \Phi, Z)
\\
\\
 {\displaystyle\frac{d\alpha}{d\tau}} =   {\displaystyle\frac{1}{P\delta_2}}\cos(\alpha)\cos(\Phi)\sin(\theta) \\
                        \hspace{8mm} -\delta_1\delta_3\bigg{(}\cos^2 {\displaystyle(\frac{\theta}{2})}\cos(\Phi+Z)-\sin^2 {\displaystyle(\frac{\theta}{2})}\cos(\Phi-Z)\bigg{)}\\
                         \hspace{8mm}+ {\displaystyle\frac{\delta_1\delta_3}{n^2}}\bigg{(}\cos(\theta)\sin(\alpha)-\cos(\alpha)\sin(\Phi)\sin(\theta)\bigg{)}F(\alpha,\Phi,Z)
\\
\\                      
 {\displaystyle\frac{d\Phi}{d\tau}} = -\delta_3 + \sin(\theta)\bigg{(}\delta_1\delta_3\cos(Z)- {\displaystyle\frac{\sin(\Phi)}{\delta_2P\sin(\alpha)}}\bigg{)} \\
 \hspace{8mm}+  {\displaystyle\frac{\delta_1\delta_3}{\tan(\alpha)}}\bigg{(}\cos^2 {\displaystyle(\frac{\theta}{2})}\sin(\Phi+Z)+\sin^2 {\displaystyle(\frac{\theta}{2})}\sin(Z-\Phi)\bigg{)}\\
 \hspace{8mm}- {\displaystyle\frac{\delta_1\delta_3}{n^2}\frac{\cos(\Phi)\sin(\theta)}{\sin(\alpha)}}F(\alpha, \Phi, Z)\\
\\
 {\displaystyle\frac{dZ}{d\tau}}=\delta_2\delta_3 P\bigg{(}\sin(\alpha)\sin(\Phi)\sin(\theta)+\cos(\alpha)\cos(\theta)\bigg{)}\\
\\
 {\displaystyle\frac{d\delta_3}{d\tau}}=- {\displaystyle\frac{\delta_1\delta_2\delta_3^3 P}{n^2}}F(\alpha, \Phi, Z)\\
\\
F(\alpha,\Phi,Z)=\sin(\alpha)\sin^2 {\displaystyle(\frac{\theta}{2})}\cos(\Phi-Z)\\
                         \hspace{22mm}+\sin(\alpha)\cos^2 {\displaystyle(\frac{\theta}{2})}\cos(\Phi+Z) +\cos(\alpha)\sin(\theta)\sin(Z).
\end{array}
\right.
\end{equation}
It is easy to observe that we can recover the classical regime by setting $\dot{\delta}_3=\dot{\gamma}/\gamma\delta_2=0$ or $F(\alpha, \Phi, Z)=0$. We now proceed to study some of the properties of the dynamical system. 

\subsection{Fixed Points.}
A common first step in the study of dynamical systems is to find and investigate the properties of fixed (stationary) points. The fixed points of the dynamical system (\ref{eq:DS}) are defined as the values in $(P, \alpha, \Phi, Z, \gamma)$, for which $(\dot{P}=\dot{\alpha}=\dot{\Phi}=\dot{Z}=\dot{\gamma}=0)$. It can be demonstrated (see Appendix 1) that the dynamical system, for $-\pi/2<\theta<0$, possesses the following values for the fixed points : 
\begin{eqnarray}
\label{eq:FP77}
P=-\gamma \tan(\theta); \hspace{10mm} \alpha =\pm\theta\pm\frac{\pi}{2}; \nonumber \\
\Phi=\pm \frac{\pi}{2}; \hspace{15mm} Z=0,\pi ; \\
\gamma=\frac{1}{\sqrt{1-{\displaystyle\frac{v_\Phi^2}{c^2}}\bigg{(}1+\tan^2(\theta)\bigg{)}}}.\nonumber
\end{eqnarray}
It is already evident from (\ref{eq:FP77}) that in the case of parallel propagation ($\theta=0$), the only fixed point is that for the trivial case $P=0$. The fixed point for the relativistic regime appears therefore similar to the classical one for parallel and oblique propagation \cite{Hamza06}. The fixed point for the relativistic regime will translate into properties found in the non-relativistic regime, but also results in different types of structures in their vicinity.  For non-zero propagation angles, fixed points identify volumes of phase-space composed of physically trapped orbits. The trapped orbits could give rise to kinetic distortions in the distribution functions, such as beams and temperature anisotropies, as was revealed in the classical non-relativistic case \cite{Osmane10}. However, because the relativistic equations possess a constraint in the form of the Lorentz factor $\gamma$, different effects are shown to arise.

\subsection{Invariants.} 
The dynamical system (\ref{eq:DS}) also possesses a number of invariants valid for the general case of oblique propagation. Knowledge of these invariants is used to construct pseudo-potential structures. In turn, these structures provide information on trapped and quasi-trapped orbits. 

\subsubsection{First invariant : $I_1$.}
Using equations (\ref{eq:norm}) to normalize equation (\ref{eq:omega_dot}), the equation describing the evolution of the gyrofrequency can be written as follow : 
\begin{equation}
\dot{\delta_3}=-\delta_3 \frac{1}{1+\frac{n^2}{\Gamma^2}}\frac{\dot{\Gamma}}{\Gamma}
\end{equation}
for $\Gamma=kp/m\omega$. Hence, this equation has an exact solution, providing the following constant of the motion :
\begin{equation}
I_1=\delta_3\sqrt{\Gamma^2+n^2}.
\end{equation}
We can write this invariant in terms of the variables $P, \alpha, \Phi, Z, $ and $\delta_3$ as follow: 
\begin{eqnarray}
I_1&=&\sqrt{\delta_3^2P^2+2\delta_3P_z/\delta_2+\delta_3^2n^2}
\end{eqnarray}
The conservation of this quantity will indicate the degree to which the constraint for the Lorentz factor (\ref{eq:constraint}) is respected in a numerical scheme.
 
\subsubsection{Second invariant : $I_2$.}
A second general invariant can be found and expressed in terms of the normalized variables as follow : 
\begin{equation}
I_2=\delta_2(n^2-1)\gamma\cos(\theta)-\delta_2P\cos(\alpha)+\delta_1\sin(\theta)\cos(Z)
\end{equation}
This invariant underlies a fundamental property of oblique propagation. One can indeed rewrite the invariant in the form $E=m\gamma c^2\sim P_\parallel$, which means that one needs a change in the parallel momentum to change the energy. This is a well-known statement resulting from the Maxwell-Lorentz invariant quantity $\mathbf{E}\cdot\mathbf{B}=0$, since a parallel component of the electric field can not be eliminated by any Lorentz translation, while the physics in a frame with $E_\parallel=0$, such as in the case of parallel propagation, is no different, therefore, than the physics in a frame where $E_\perp=E_\parallel=0$ for which energy is a constant of the motion. 

\section{Special Cases.}
\subsection{Parallel Propagation : $\theta=0$.}
The wave-particle interaction problem has overwhelmingly been treated for the special case of parallel propagation. Even though we do not present any new result in this section, we find it useful to briefly discuss the parallel case as a means of comparison to the general oblique case. Setting $\theta=0$ in (\ref{eq:DS}), we recover the following dynamical system : 

\begin{equation}
\label{eq:DS_0}
\left\{ 
\begin{array}{l l} 
{\displaystyle\frac{dP}{d\tau}}= - {\displaystyle\frac{\delta_1\delta_3 P}{n^2}}\cos(\alpha)F(\alpha, \Phi, Z)
\\
\\
 {\displaystyle\frac{d\alpha}{d\tau}} =  -\delta_1\delta_3\cos(\Phi+Z)+{\displaystyle\frac{\delta_1\delta_3}{n^2}}\sin(\alpha)F(\alpha,\Phi,Z)
\\
\\                      
 {\displaystyle\frac{d\Phi}{d\tau}} = -\delta_3+{\displaystyle\frac{\delta_1\delta_3}{\tan(\alpha)}}\sin(\Phi+Z)\\
\\
 {\displaystyle\frac{dZ}{d\tau}}=\delta_2\delta_3 P\cos(\alpha)\\
\\
 {\displaystyle\frac{d\delta_3}{d\tau}}=- {\displaystyle\frac{\delta_1\delta_2\delta_3^3 P}{n^2}}F(\alpha, \Phi, Z)\\
\\
F(\alpha,\Phi,Z)=\sin(\alpha)\cos(\Phi+Z).
\end{array}
\right.
\end{equation}
In addition to the two invariants $I_1$ and $I_2$, equations (\ref{eq:DS_0}) also possesses the following constant of motion\footnote[22]{This invariant is the relativistic equivalent found in previous studies, i.e., \cite{Sudan66} and \cite{Hamza06}.} for $n^2\ne 1$ : 
\begin{equation}
\label{eq:parallel_invariant}
(\delta_2P\cos(\alpha)-1)^2=2\delta_1\delta_2\frac{n^2-1}{n^2}P\sin(\alpha)\sin(\Phi+Z).
\end{equation}\\
Moreover, the existence of physically trapped orbits for $\theta=0$ requires that $\cos(\alpha)=\cos(\Phi+Z)=0$, hence, $\alpha=\Phi+Z=\pi/2$. However, this conditions results in $\dot{\Phi}\ne 0$. Aside from the trivial case of $P=0$, no fixed point exists and the parallel propagation has the particular distinction, with respect to oblique propagation, to not possess solutions for which a particle could be trapped in $Z$.\\
The parallel case has been studied in both the classical and relativistic regime. The classical treatment covered by \textit{Matsumoto}\citep[][]{Matsumoto85} and  \textit{Hamza et al.}\citep[][]{Hamza06}, have shown that one can find exact solutions in terms of elliptical integrals. \textit{Lutomirski and Sudan}\citep{Sudan66} have studied the relativistic case showing that similar solutions were also possible.  \textit{Roberts and Buchsbaum} \citep[][]{Roberts64} have also treated the relativistic case with a special focus on the case $n^2=1$, for which a  cyclotron-resonant particle was shown to gain energy indefinitely, while for $n^2\ne 1$, the particle simply becomes phase trapped at cyclotron-resonance with no net gain in energy on average. Using the invariant in equation (\ref{eq:parallel_invariant}) as well as $I_1$ and $I_2$, those results can be expressed in terms of a pseudo-potential equation in the parallel component of the momentum that we write as  $y=\delta_2P\cos(\alpha)=\delta_2 P_\parallel$ : 
\begin{eqnarray}
\label{eq:potential_parallel}
\frac{\dot{y}^2}{2}&=&-V(y; \delta_1,\delta_2, n^2, I_2)\nonumber\\
	&=&\frac{\delta_1^2}{2}\Bigg{[}\frac{n^2-1}{n^2}\Bigg{]}^2\Bigg{[}\frac{n^2-1}{\sigma(y)^2}-2\frac{y}{\sigma(y)}-\bigg{(}y^2+n^2\delta_2\bigg{)}\Bigg{]}-\frac{1}{8}\sigma(y)^2\bigg{(}y-1\bigg{)}^4
\end{eqnarray}
for the function $\sigma(y)=(n^2-1)/(I_2+y)$. Solutions to equation (\ref{eq:potential_parallel}) for $V(y ; \delta_1, \delta_2, n^2, I_2) < 0$ are bound states of the system for as far as parallel momentum is concerned. However, this only holds for $n^2\ne 1$. In the case of $n^2=1$, we can easily recover the unlimited acceleration found by \textit{Roberts and Buchsbaum}\citep[][]{Roberts64}  from the invariants of the motion. Setting $\theta=0$ and $n^2=1$ for $I_2$, one finds that $P_\parallel$ is constant. That is, if a particle is at cyclotron-resonance, it will remain so forever (or until the wave damps), and gain energy indefinitely. It is demonstrated in the remainder of the report, that unlimited acceleration is also possible for oblique propagation and that it underlies a specific property of the fixed points.\\

\subsection{Perpendicular Propagation : $\theta=-\pi/2$.}
We now investigate the purely perpendicular case, as its treatment will be useful to characterize the dynamics for propagation angles that increase towards $|\pi/2|$. The dynamical system is written in the following form : 
\begin{equation}
\label{eq:DS_90}
\left\{ 
\begin{array}{l l} 
{\displaystyle\frac{dP}{d\tau}}= -{\displaystyle\frac{\sin(\alpha)\cos(\Phi)}{\delta_2}}+{\displaystyle\frac{\delta_1\delta_3 P}{n^2}}\sin(\alpha)\sin(\Phi)F(\alpha, \Phi, Z)
\\
\\
 {\displaystyle\frac{d\alpha}{d\tau}} = -{\displaystyle\frac{1}{\delta_2P}}\cos(\alpha)\cos(\Phi)+\delta_1\delta_3\sin(\Phi)\sin(Z)\\
 \hspace{6mm}+{\displaystyle\frac{\delta_1\delta_3}{n^2}}\cos(\alpha)\sin(\Phi)F(\alpha,\Phi,Z)
\\
\\                      
 {\displaystyle\frac{d\Phi}{d\tau}} = -\delta_3-\delta_1\delta_3\cos(Z)+{\displaystyle\frac{\sin(\Phi)}{\delta_2P\sin(\alpha)}}\\
 \hspace{6mm}+{\displaystyle\frac{\delta_1\delta_3}{\tan(\alpha)}}\cos(\Phi+Z)+{\displaystyle\frac{\delta_1\delta_3\cos(\Phi)}{n^2\sin(\alpha)}}\cos(\Phi+Z)F(\alpha, \Phi, Z)\\
\\
 {\displaystyle\frac{dZ}{d\tau}}=-\delta_2\delta_3 P\sin(\alpha)\sin(\Phi)\\
\\
 {\displaystyle\frac{d\delta_3}{d\tau}}=- {\displaystyle\frac{\delta_1\delta_2\delta_3^3 P}{n^2}}F(\alpha, \Phi, Z)\\
\\
F(\alpha,\Phi,Z)=\sin(\alpha)\cos(\Phi)\cos(Z)-\cos(\alpha)\sin(Z).
\end{array}
\right.
\end{equation}

Similarly to the parallel case, the dynamical system (\ref{eq:DS_90}) possesses its own set of invariants written as : 
\begin{equation}
\left\{
\begin{array}{l l} 
I_4=\delta_1\cos(Z)+\delta_2P\cos(\alpha) \nonumber\\
I_5=\delta_2P\cos(\Phi)\sin(\alpha)+\delta_1\sin(Z)+Z+\tau . \nonumber
\end{array}
\right.
\end{equation}
With the fixed point analysis for this particular case showing that no fixed points exists, i.e., a particle can not be physically trapped, we make the assumption that the solution for $Z$ takes the form of a linear relationship in time : $Z=Z_0+\beta \tau$, with $Z_0$ as the initial condition and $\beta$ as a constant. Replacing the solution for $Z$ in the invariant $I_5$ results in the following expression : 
\begin{equation}
I_5=\delta_2P\cos(\Phi)\sin(\alpha)+\delta_1\sin(Z_0+\beta \tau)+Z_0+\beta \tau+\tau
\end{equation}
It is therefore evident that for $\dot{I_5}=0$ to be true, the term $(\beta+1)\tau$ must be either zero, or compensated by  the momentum in $\hat{x}$, $P_x=P\sin(\alpha)\cos(\Phi)$, to grow to minus infinity as $\tau$ goes to infinity. In the absence of accessible Landau and cyclotron resonances, the latter solution does not appear acceptable. We can qualitatively demonstrate this assumption by noting that for $\tau \gg \delta_1$, the following approximation must be respected : $\frac{\gamma}{\tan(\Phi)}\simeq\frac{1-\beta}{\delta_2\beta}\tau$. Hence, either a) $\gamma\rightarrow\infty$, or b) $\Phi\rightarrow 0$.In the first case, if $\gamma\rightarrow\infty$, then $P\rightarrow\infty$ as well. Hence for $I_4$ to be constant, stationary solutions giving $P_y=P\cos(\alpha)\sim constant$ are required. Such solutions would necessitate $\dot{P_y}\sim0$. Such constraint means that either $P_z=0$ or $Z=0$. But both solutions are unacceptable since they would imply the existence of a fixed point, which has been demonstrated to not exist for the special case of perpendicular propagation. In the second case, the requirement that $\Phi\rightarrow 0$ means that since $\delta_3P\simeq v\leq c$ is bounded, $\dot{Z}\rightarrow 0$, which is in contradiction with the evidence that $Z$ must be linear in time because of a zero parallel electric field. We are therefore left with the assumption that $\beta\sim-1$, an assumption that can indeed be verified by numerical integration. \\
Without any loss of generality, we set $Z_0=0$, resulting in the solution $Z=-\tau$. Using $I_4$ we find the following solutions for $P_\parallel$ : 
\begin{equation}
\delta_2P_\parallel=I_4-\delta_1\cos(\tau).
\end{equation}
Similarly, the solution for $P_x$ can be directly found from $I_5$ : 
\begin{equation}
\delta_2P_x=I_5+\delta_1\sin(\tau).
\end{equation}
Using those two solutions we can find the exact differential for $\delta_3$ : 
\begin{equation}
\frac{d\delta_3}{\delta_3^3}=-\frac{\delta_1}{n^2}[I_5\cos(\tau)+I_4\sin(\tau)]d\tau.
\end{equation}
Hence, the following solutions for $\delta_3$ : 
\begin{equation}
\label{eq:gamma_90}
\frac{1}{\delta_3^2}-{\frac{1}{\delta_3^2(0)}=\frac{2\delta_1}{n^2}[I_5\sin(\tau)+I_4-I_4\cos(\tau)]}.
\end{equation}
We can finally find the exact solution for the last variable in terms of $\tau$ from the dynamical system equation in $Z$, that is : 
\begin{equation}
\delta_2P_z=\sqrt{\frac{1}{\delta_3^2(0)}+\frac{2\delta_1}{n^2}[I_5\sin(\tau)+I_4-I_4\cos(\tau)]}.
\end{equation}
We have therefore derived exact solutions for the perpendicular case, based on the existence of two invariants and the nonexistence of fixed points. Two limiting cases can be deduced from these solutions.  If the wave is sustained for long periods, such that the time of interaction with the particles $\tau_{int}\sim 1/\epsilon\omega$ for $\epsilon \ll 1$, the perpendicular propagation results in phase trapped orbits with no net gain of energy on average. In the opposite case where the interaction would be short-lived such that $\tau_{int}\sim\epsilon/\omega$, we can calculate the average increment in energy during the time of interaction. If we write (\ref{eq:gamma_90}) in terms of $E=m\gamma c^2$, and assume large amplitude, low-frequency waves such that $\delta_1/\delta_2 \sim 1$, then $E/E_0=\sqrt{1+2\delta_1^2/\delta_2^2n^2\gamma_0^2}\sim 1+v_\Phi^2/c^2$  and a particle gains energy of the order of $\Delta E/E\sim v_\Phi^2/c^2$ for every interaction. Given a prescription in the probability of interaction $P(v_\Phi, \Delta t)$ with an electromagnetic wave of phase-speed $v_\Phi$, one could build a map to describe the nonlinear interaction of a particle in a relativistic turbulent plasma composed of highly oblique electromagnetic waves. This qualitative analysis for purely perpendicular wave applies for particles that do not belong to Landau or cyclotron resonance.  

\section{Cyclotron and Landau Resonances.}
\subsection{Stochastic acceleration at Cyclotron resonance : $\omega-k_\parallel v_\parallel=\pm s\Omega_0/\gamma$.}
\begin{figure}[ht] %  figure placement: here, top, bottom, or page
   \centering
   \includegraphics[width=0.50\textwidth]{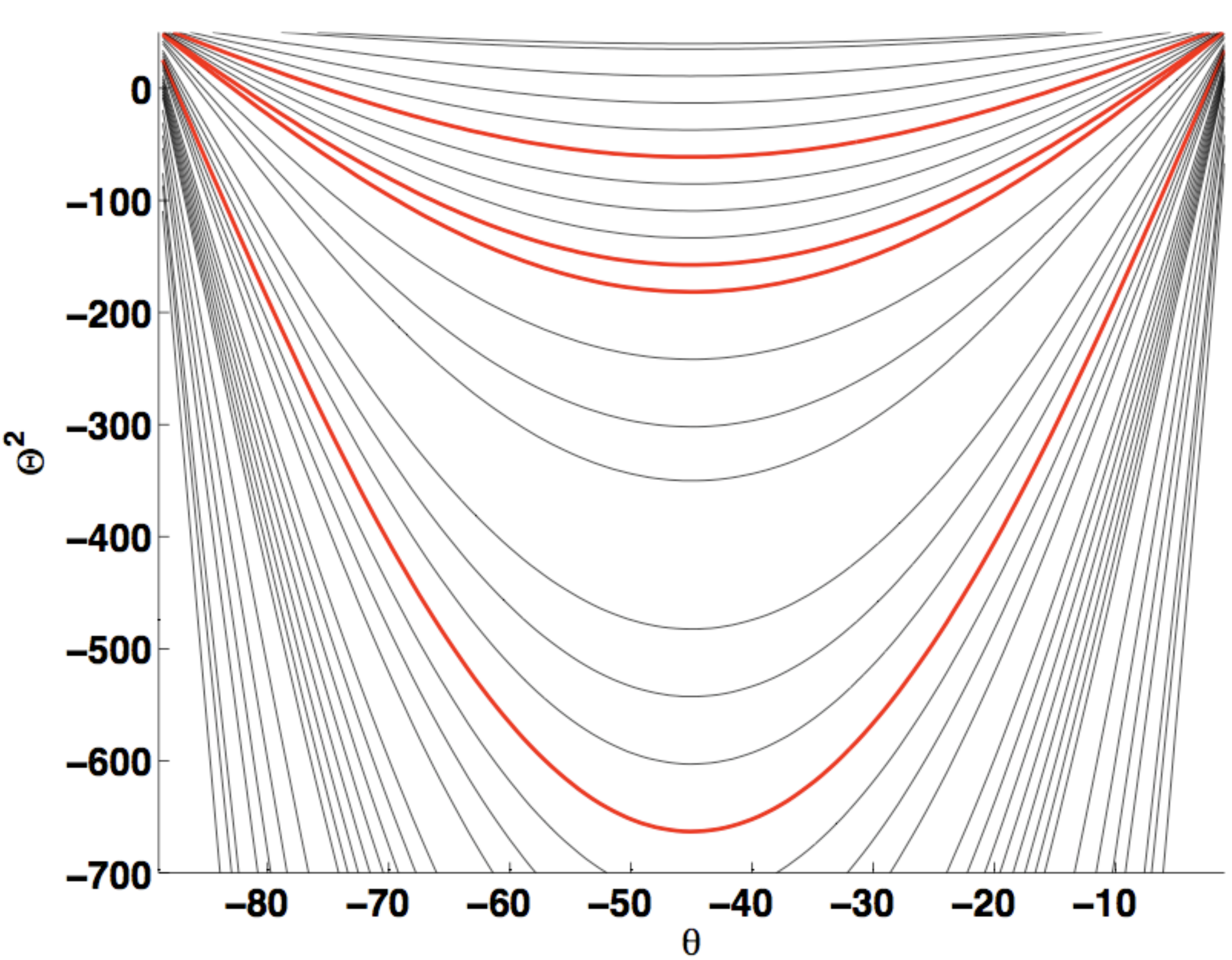}%{nsd_template_fig}
   \caption{Squared frequency $\Theta^2$ as a function of the propagation angle $\theta$ for a relativistic particle in cyclotron resonance. Each curves are for different values in the wave-amplitude parameter spanning $0.01\le\delta_1\le1$. The four bold red lines correspond, from top to bottom, to $\delta_1=(0.05, 0.09, 0.1, 0.3)$}
   \label{fig1}
\end{figure}

\begin{figure}[p] %  figure placement: here, top, bottom, or page
   \centering
   \includegraphics[width=0.5\textwidth]{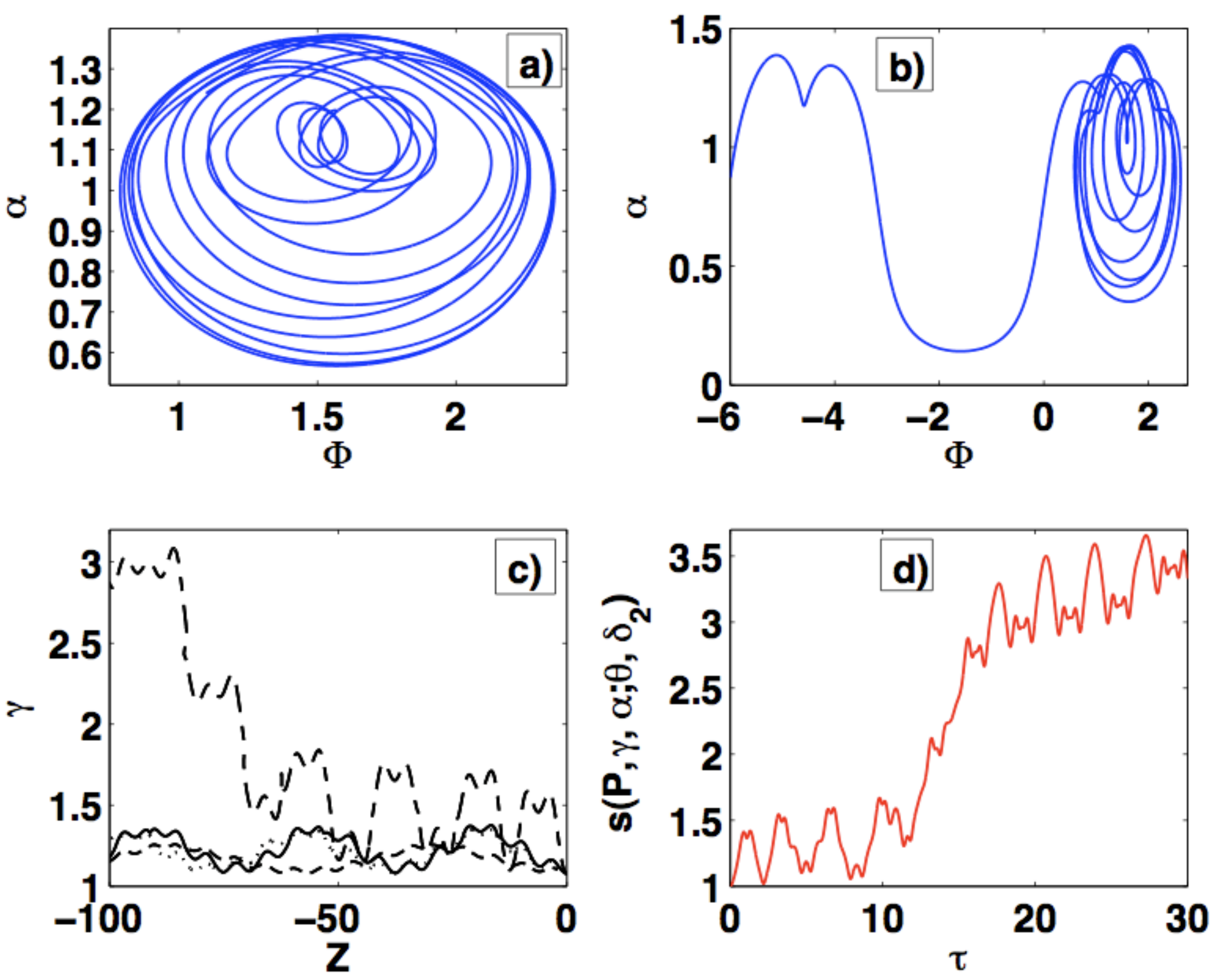}%{nsd_template_fig}
   \caption{a)Pitch angle $\alpha$ vs dynamical gyrophase $\Phi$ for $\Theta^2>0$. The particle is phase-trapped. b)Pitch angle $\alpha$ vs dynamical gyrophase $\Phi$ for $\Theta^2<0$. The particle is quasi-trapped in phase-space c) Lorentz factor $\gamma$ vs Z for  $\delta_1=(0.05, 0.09, 0.1, 0.3)$. When $\Theta^2<0$, the orbit is unstable in $\gamma$ and depart from the forced harmonic oscillation observed for $\Theta^2>0$. d)Resonance condition quantified by $s(P, \gamma, \alpha ; \theta, \delta_2)$ for the case of $\Theta^2<0$. The particle travels through multiple resonances as it gains energy through repeated kicks.}
   \label{fig2}
\end{figure}

\begin{figure}[ht] %  figure placement: here, top, bottom, or page
   \centering
   \includegraphics[width=0.50\textwidth]{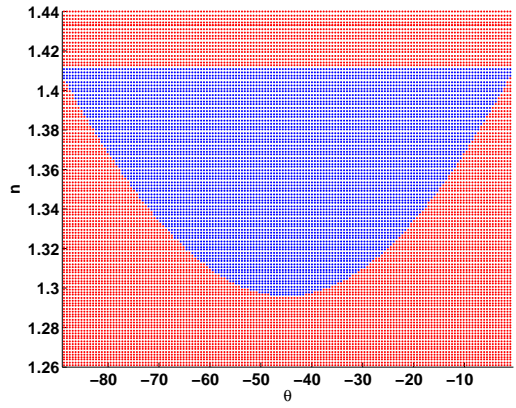}%{nsd_template_fig}
   \caption{Arnold tongue in the parameter space $(\theta, n=c/v_\Phi$), for $\delta_1=0.3$, $\delta_2=1$.}
   \label{FIG6}
\end{figure}

\begin{figure}[ht] %  figure placement: here, top, bottom, or page
   \centering
   \includegraphics[width=0.50\textwidth]{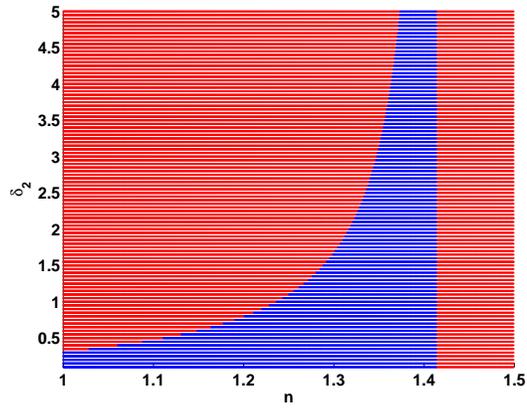}%{nsd_template_fig}
   \caption{Arnold tongue in the parameter space $(n, \delta_2$), for $\delta_1=0.5$, $\theta=45^o$. }
   \label{FIG7}
\end{figure}

The most commonly studied problems of wave-particle interactions have been addressed in the context of cyclotron-resonance. 
However, we demonstrate below that the case of cyclotron-resonance contains further intricacies when the general case of oblique propagation and nonlinear interaction is treated in the relativistic limit. In order to do so, we construct a pseudo-potential function for a particle crossing resonances.\\
The resonance condition is written in terms of the normalized variables as : 
\begin{equation}
\gamma\sin^2(\theta)-P\cos(\alpha)\cos(\theta)=\pm s/\delta_2.
\end{equation}
Using the resonance condition to replace the expression of $P\cos(\alpha)$ in $I_2$, we find the following expression : 
\begin{equation}
\label{eq:invariant_cyclotron}
I_2\cos(\theta)\mp s=\delta_2\gamma(n^2\cos^2(\theta)-1)+\delta_1\cos(Z)\sin(\theta)\cos(\theta).
\end{equation}
If $\gamma$ and $Z$ do not have singularities in their derivatives when the resonance condition is respected,  the following relationship must be satisfied : 
\begin{equation}
\frac{d\gamma}{d\tau}=\frac{\delta_1\sin(\theta)\cos(\theta)}{\delta_2(n^2\cos^2(\theta)-1)}\sin(Z)\frac{dZ}{d\tau}.
\end{equation}
We can find an expression between $\dot{Z}$ and $\gamma$ from the invariant $I_1$. In order to do so, we write the invariant quantity in the following form : 
\begin{eqnarray}
\label{eq:approxZ}
\frac{dZ}{d\tau}-\frac{n^2-1}{2} & = &-\frac{P^2+n^2}{2\gamma^2}  \nonumber \\
				&=&-\frac{n^2}{2\gamma^2}\Bigg{[}\bigg{(}\frac{P}{n}+1\bigg{)}^2-2\frac{P}{n}\Bigg{]}\nonumber \\
				&\simeq&-\frac{n^2}{2\gamma^2}; \hspace{49mm}{if \hspace{2mm}\frac{P}{n}\ll 1}. 
\end{eqnarray}
Hence, using equation (\ref{eq:approxZ}) in addition to (\ref{eq:invariant_cyclotron}) we can  replace $\dot{Z}$ and $\sin(Z)$ and  find a pseudo-potential equation in $\gamma$  of the form : 
\begin{equation}
\label{eq:potential_gamma}
\frac{\dot{\gamma}^2}{2}+V(\gamma ; \delta_1, \delta_2, \theta)=0
\end{equation}
for a pseudo-potential written as : 
\begin{eqnarray}
V(\gamma ; \delta_1, \delta_2, \theta)&=&-\frac{1}{2}\Bigg{[}\frac{n^2-1}{2}-\frac{n^2}{2\gamma^2}\Bigg{]}^2\nonumber \\
        &\times& \Bigg{[}\Bigg{(}\frac{\delta_1}{\delta_2}\frac{\sin(\theta)\cos(\theta)}{n^2\cos^2(\theta)-1}\Bigg{)}^2-\Bigg{(}\frac{I_2\cos(\theta)\mp s}{\delta_2n^2\cos^2(\theta)-\delta_2}-\gamma}{\Bigg{)}^2\Bigg{]} \nonumber\\
        &=&-\frac{1}{8}\Bigg{[}\beta_1-\frac{\beta_1+1}{\gamma^2}\Bigg{]}^2\Bigg{[}\beta_2^2-\Bigg{(}\beta_3-\gamma\Bigg{)}^2\Bigg{]},
\end{eqnarray}
for the set of constants $\beta_1, \beta_2, \beta_3$ defined as follows : 
\begin{equation}
\beta_1=n^2-1\nonumber
\end{equation}
\begin{equation}
\beta_2=\frac{\delta_1}{\delta_2}\frac{\sin(\theta)\cos(\theta)}{n^2\cos^2(\theta)-1}\nonumber
\end{equation}
\begin{equation}
\beta_3=\frac{I_2\cos(\theta)\mp s}{\delta_2 n^2\cos^2(\theta)-\delta_2}
\end{equation}
If we set the initial conditions $Z_0=0$ and $\gamma_0=1$, we can write $\beta_3=\beta_2+1$. Taking the second derivative of (\ref{eq:potential_gamma}), we find the following expression : 
\begin{eqnarray}
\ddot{\gamma}&=&\frac{1}{4}\beta_1^2\gamma-\frac{1}{4}\beta_2\beta_1^2+\frac{1}{2}\frac{\beta_3\beta_1(\beta_1+1)}{\gamma^2}+\frac{1}{2}\frac{(\beta_1+1)^2+(\beta_3^2-\beta_2^2)(\beta_1+1)}{\gamma^3}\nonumber \\
&+&\frac{3}{4}\frac{\beta_3(\beta_1+1)^2}{\gamma^4}-\frac{1}{2}\frac{(\beta_2^2-\beta_3^2)(\beta_1+1)^2}{\gamma^5}.
\end{eqnarray}
This equation can be used to treat the cyclotron-resonance for different limits. We hereafter focus on the relativistic low energy case for which $\gamma=\gamma_0+\delta\gamma$, with $\delta\gamma\ll\gamma_0$. Using Newton's approximation to express $\gamma^{-n}\simeq\gamma_0^{-n}(1-n\delta\gamma/\gamma_0)$ and setting $\gamma_0=1$, we find the following forced oscillator equation : 
\begin{equation}
\ddot{\delta\gamma}+\Theta^2\delta\gamma=\Lambda(\beta_1, \beta_2),
\end{equation}
for the frequency squared : 
\begin{eqnarray}
\Theta^2 &=&-\frac{1}{4}\beta_1^2-\beta_3\beta_1(\beta_1+1)+\frac{3}{2}[(\beta_1+1)^2+(\beta_3^2-\beta_2^2)(\beta_1+1)]\nonumber \\
&+& 3\beta_3(\beta_1+1)^2-\frac{5}{2}[(\beta_2^2-\beta_3^2)(\beta_1+1)^2], 
\end{eqnarray}
and the constant forcing term 
\begin{eqnarray}
\Lambda&=&-\frac{1}{4}\beta_1^2(\beta_3-1)+\frac{1}{4}\beta_3(\beta_1+1)(\beta_1+3)\nonumber \\
&+& \frac{1}{2}[(\beta_1+1^2)+(\beta_3^2-\beta_2^2)(\beta_1+1)]-\frac{1}{2}(\beta_2^2-\beta_3^2)(\beta_1+1)^2.
\end{eqnarray}
Figure \ref{fig1} represents the dependence of $\Theta$ as a function of $\theta$ for fixed values of $\delta_1$. It is clear that for the range of chosen parameters ($v_\Phi/c\sim .70, \delta_2=1$), the oscillations in $\delta\gamma$ can evolve from harmonic solutions to hyperbolic solutions as the amplitude of the wave increases. As a result of a large wave-amplitude, that is $\delta_1$ growing, a wide range of propagation angles will result in hyperbolic perturbations for a relativistic particle in cyclotron resonance.  Figure \ref{fig2} represents the transition from $\Theta^2>0$ to $\Theta^2<0$. As the wave-amplitude increases, the particle transit from trapped-orbits to quasi-trapped orbits in phase-space. If the amplitude is further increased, the orbit becomes stochastic. Quasi-trapped and stochastic orbits are resulting from the wandering of the particle from one cyclotron harmonic to another. Hence, the particle gains energy stochastically. This result is an extension of the overlapping resonances studied by  \textit{Smith and Kaufman}\citep[][]{Smith78} for classical regimes. Using a restrictive choice of parameters, they found that wave amplitudes of the order of $\delta_1=\delta B/B_0\geq 15$ were necessary to have overlapping resonances. However, our analysis shows that there is a window in parameter space belonging to the relativistic regime that allows for the overlapping of resonances for amplitudes two orders of magnitude smaller. Similarly to the classical case, large-amplitudes translate into a broadening of the phase-trapping cell. Trapping cells are also largest for propagations at $\theta=45^o$. Plotted in figures (\ref{FIG6}) and (\ref{FIG7}) are Arnold tongues, that is regions of parameter space ($n^2, \theta, \delta_1, \delta_2$) leading to stochastic orbits. It is evident from the Arnold tongues that even though the effect described by our analysis is purely relativistic, a wide range of parameters can result in stochastic orbits.\\ 
It should be noted that even though the equations presented in this section also apply to the case of Landau resonance, the parameter space, in which unstable orbits and overlapping can operate, belongs to velocities that must go beyond the speed of light. Therefore, the aforementioned result applies specifically to the case of cyclotron-resonances. 


%\subsection{Particle Confined along $\mathbf{\it{k}}$ : $\dot{Z}=Z=0$.}
%We are now interested with the special case of a particle confined along the propagation vector. The wave-particle interaction with a particle constrained physically by a force $f_z$, has interesting applications in the context of shock, and more specifically to the problem of the injection and acceleration at relativistic shocks. The dynamical system for such a case writes as follow : 
%\begin{equation}
%\label{eq:DS}
%\left\{ 
%\begin{array}{l l} 
%{\displaystyle\frac{dP}{d\tau}}=\sin(\alpha)\cos(\Phi)\sin(\theta)/\delta_2 \\
%\\
% {\displaystyle\frac{d\alpha}{d\tau}} =   {\displaystyle\frac{1}{P\delta_2}}\cos(\alpha)\cos(\Phi)\sin(\theta) \\
%                         \hspace{8mm} -\delta_1\delta_3\cos(\Phi)\cos(\theta)+ {\displaystyle\frac{\delta_1\delta_3}{n^2}}\cos(\theta)\cos(\Phi)
%\\
%\\                      
% {\displaystyle\frac{d\Phi}{d\tau}} = -\delta_3 + \sin(\theta)\bigg{(}\delta_1\delta_3- {\displaystyle\frac{\sin(\Phi)}{\delta_2P\sin(\alpha)}}\bigg{)} \\
% \hspace{8mm}+  {\displaystyle\frac{\delta_1\delta_3}{\tan(\alpha)}}\cos(\theta)\sin(\Phi)-{\displaystyle\frac{\delta_1\delta_3}{n^2}\cos^2(\Phi)\sin(\theta)}\\
%\\
% {\displaystyle\frac{dZ}{d\tau}}=\delta_2\delta_3 P\bigg{(}\sin(\alpha)\sin(\Phi)\sin(\theta)+\cos(\alpha)\cos(\theta)\bigg{)}=0\\
%\\
% {\displaystyle\frac{d\delta_3}{d\tau}}=- {\displaystyle\frac{\delta_1\delta_2\delta_3^3 P}{n^2}}F(\alpha, \Phi, Z)\\
%\\
%F(\alpha,\Phi,Z=0)=\sin(\alpha)\cos(\Phi).
%\end{array}
%\right.
%\end{equation}
\begin{figure}[ht] %  figure placement: here, top, bottom, or page
   \centering
   \includegraphics[width=0.50\textwidth]{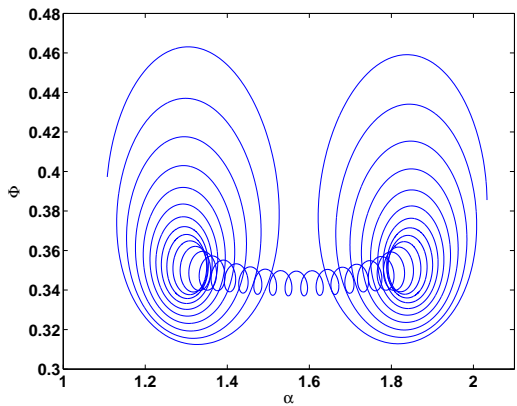}%{nsd_template_fig}
   \caption{Case $\theta<\theta_c=\arctan{\sqrt{n^2-1}}$. Particle orbit for parameters $\delta_1=0.1$, $\delta_2=0.0696$, $n^2=4$, $\theta=\theta_c-1^o$ and initial conditions $v_{x0}'=0$, $v_{y0}'=-v_\Phi\tan(\theta)-1.6v_\Phi$, $v_{z0}'=-v_\Phi$, $Z_0=0$. The particle is physically and phase trapped. It bounces back and forth in the potential well with no net gain in energy.}
   \label{fig3}
\end{figure}

\begin{figure}[ht] %  figure placement: here, top, bottom, or page
   \centering
   \includegraphics[width=0.50\textwidth]{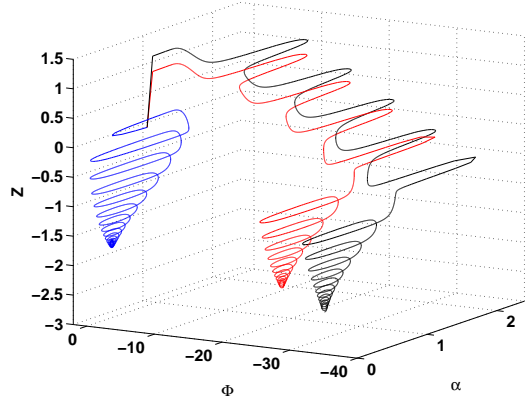}%{nsd_template_fig}
   \caption{ Case $\theta=\theta_c$. Three particle orbits seeded with different initial conditions show that the attractor is periodic. Parameters $\delta_1=0.1$, $\delta_2=0.0696$, $n^2=4$. The orbit is locked in pitch-angle $\alpha$ and dynamical gyrophase $\Phi$, trapped along $Z$ and accelerated uniformly.}
   \label{fig4_2}
\end{figure}

\begin{figure}[ht] %  figure placement: here, top, bottom, or page
   \centering
   \includegraphics[width=0.50\textwidth]{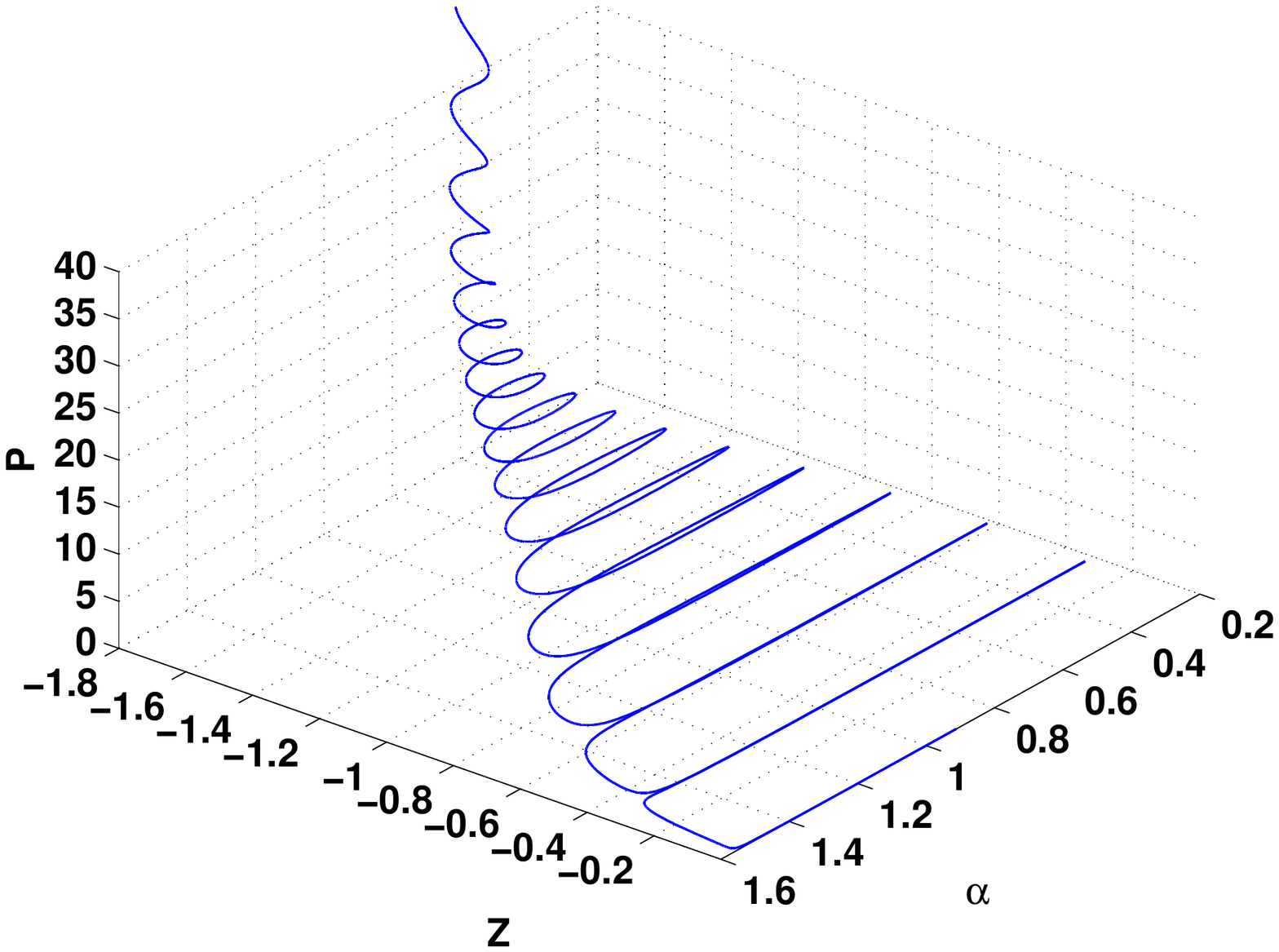}%{nsd_template_fig}
   \caption{ Case $\theta=\theta_c$. Parameters $\delta_1=0.1$, $\delta_2=0.0696$, $n^2=4$. The orbit is locked in pitch-angle $\alpha$ and dynamical gyrophase $\Phi$, trapped along $Z$ and accelerated uniformly.}
   \label{fig4_1}
\end{figure}

\subsection{Hopf bifurcation at Landau resonance : $\omega=k_\parallel v_\parallel$.}
\begin{figure}[ht] %  figure placement: here, top, bottom, or page
   \centering
   \includegraphics[width=0.50\textwidth]{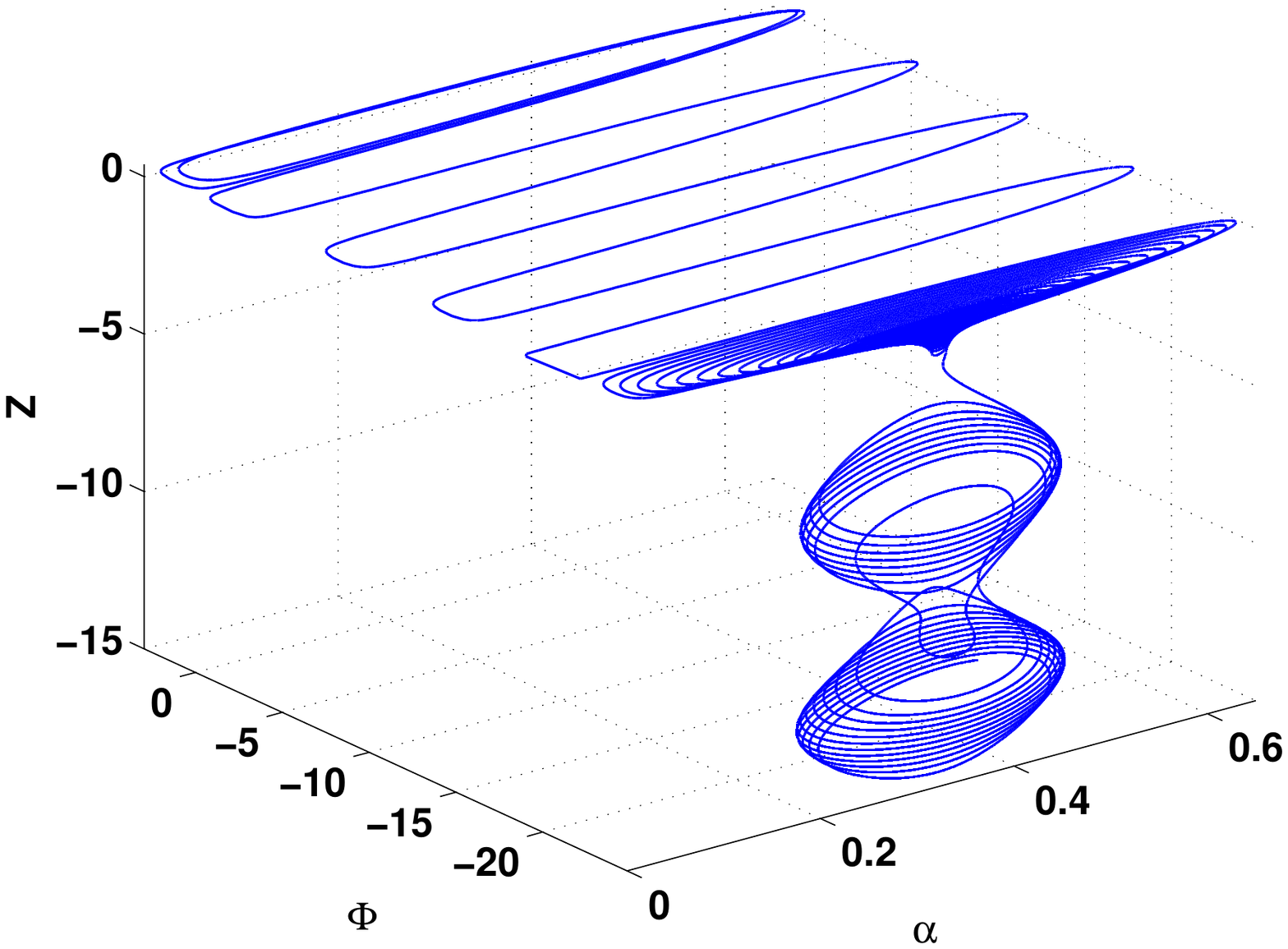}%{nsd_template_fig}
   \caption{ Case $\theta>\theta_c$.Particle orbits for parameters $\delta_1=0.1$, $\delta_2=0.0696$, $n^2=9$. The attractor is lost and can give rise to quasi-trapped orbits in the dynamical phase angles.}
   \label{fig5_1}
\end{figure}

\begin{figure}[ht] %  figure placement: here, top, bottom, or page
   \centering
   \includegraphics[width=0.50\textwidth]{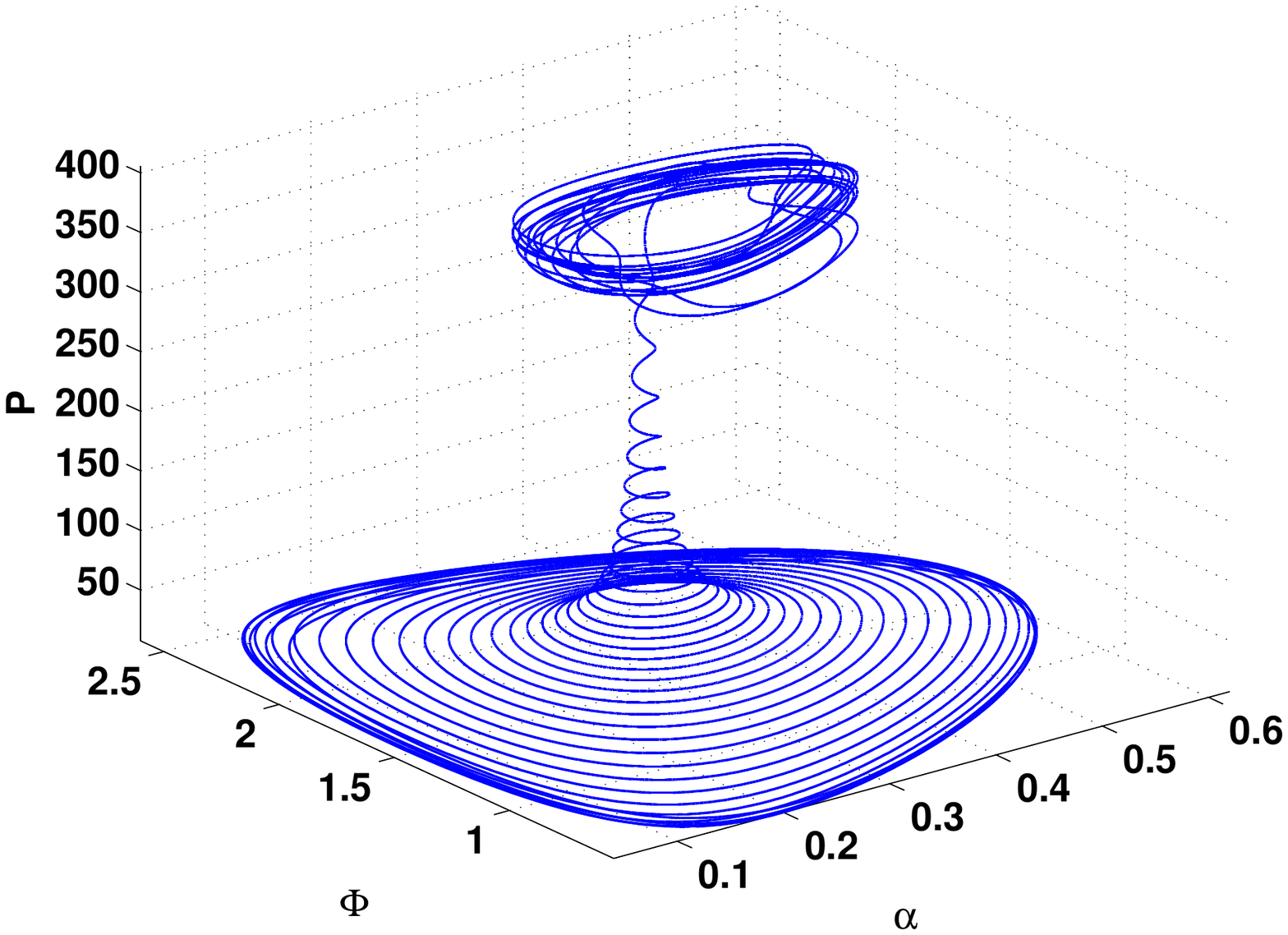}%{nsd_template_fig}
   \caption{ Case $\theta>\theta_c$.Particle orbits for parameters $\delta_1=0.1$, $\delta_2=0.0696$, $n^2=9$. Despite the lose of the attractor particles can still be accelerated to relativistic energy levels.}
   \label{fig5_2}
\end{figure}

A fundamental property of a given dynamical system can be deduced by investigating whether the phase-space density and volume is conserved or not. That is, whether or not Liouville's theorem applies \cite{Regev}. The validity of Liouville's theorem provides the possibility to construct distribution functions and follow their evolution in time. Non-conservation of phase-space density, either locally or globally, stems from the existence of either attractors or non-bounded orbits. Making use of the invariant $I_2$, we compute the divergence of the flow in phase-space as follow : 
\begin{eqnarray}
\frac{1}{V}\frac{dV}{dt}&=&\vec{\nabla}\cdot \frac{d\vec{\xi}}{dt}\nonumber \\
	&=&\frac{\partial \dot{P}_x}{\partial P_x} +\frac{\partial \dot{P}_y}{\partial P_y} +\frac{\partial \dot{P}_z}{\partial P_z} +\frac{\partial \dot{Z}}{\partial Z} \nonumber \\
	&=&-\frac{\dot{\gamma}}{\gamma},
	\label{Liouville} 
\end{eqnarray}
for the volume in phase-space $V$ and the phase-space vector coordinate $\vec{\xi}=(P_x, P_y, P_z, Z)$. Hence, equation (\ref{Liouville}) hints at the existence of an attractor if the volume in phase-space shrinks as $\gamma \rightarrow \infty$. In the case where the particle's energy oscillates back and forth such as for a volume of physically trapped orbits, we can consider Liouville's theorem to apply. But it can be shown that such an attractor does exist. A recently published Letter has shown that the attractor arises from a change in parameters that results in the bifurcation of the orbits around the fixed points \cite{Osmane12}. Indeed, the stability analysis in Appendix B demonstrates that to every fixed points, combined values in $(\theta, n^2)$, satisfying the condition $n^2-1=\tan^2(\theta)$, correspond to a bifurcation in stability \footnote[11]{The fixed point at $Z=\pi$ is unstable. Hence, no physical trapping is possible for $\theta<\theta_c$ and no uniform acceleration can arise for $\theta=\theta_c$.}. That is, an orbit, close to the fixed point will experience a transition from a (marginally) stable orbit to an unstable orbit.  We observe that when the condition in parameter space is respected, and for a large enough amplitude of the wave magnetic field, the real part of one of the eigenvalues becomes positive. This type of bifurcation for pairs of complex conjugate eigenvalues crossing through the imaginary axis, is the well-known Hopf bifurcations\cite{Gucken}. \\
Represented in Figures (\ref{fig3}), (\ref{fig4_2}), (\ref{fig4_1}),  (\ref{fig5_1}) and (\ref{fig5_2}), are typical families of orbits for parameters below, equal to, and above the propagation angle at the Hopf bifurcations ($\theta_c=\arctan{\sqrt{n^2-1}}$) for a given refractive index $n$, respectively. The wave parameters are chosen for a large-amplitude ($\delta_1=0.1$), low-frequency wave ($\delta_2=0.0696$), but similar results also apply to frequencies of the order of the gyrofrequency as long as the wave-amplitude is sufficiently large to allow physical trapping. We can observe that when $\theta<\theta_c$, the particle becomes physically and phase-trapped in the phase-space region centered at the fixed point. The particle eventually closes unto itself with no net gain on average in energy.  For $\theta=\theta_c$, the particles belonging to the basin of attraction centered around the fixed point becomes locked in pitch-angle $\alpha$ and dynamical gyrophase $\Phi$, and trapped along $Z$. This locking effect results in the divergence of the momentum to infinity under a uniform acceleration. This mechanism is similar to the surfatron process commonly studied in the physics of lasers and in the problem of wave-particle acceleration in astrophysical shocks\cite{Katsouleas, Karimabadi90, Chernikov92}. Such an effect is purely relativistic and requires the presence of the Lorentz-invariant parallel electric field. The violation of Liouville's theorem belongs to volumes composed of these surfing and trapped orbits. However, since the surfing acceleration is so efficient, a wave would be expected to damp away before considerations for self-consistency and collisions are deemed necessary. The case of $\theta>\theta_c$ manifests itself through the loss of stability of the fixed point and the evolution of the attractor into two-dimensional tori. The particle is initially trapped in the $\alpha$, $\Phi$ and $Z$ plane, but eventually becomes untrapped in $Z$ while its orbit never closes. Such a regime in parameter space can as well result in the acceleration of particles. Figure \ref{fig4_1} shows that despite the incapacity to trap physically the orbits, the particle can be accelerated to relativistic levels. It is therefore clear from the above examples that the fixed points manifest themselves differently as a function of the wave obliquity and that the propagation angle is a critical parameter for relativistic orbits in the presence of large-amplitude waves.

\section{Discussion.}
\subsection{General framework}
A general framework for understanding the wave-particle interaction for a monochromatic wave can be drawn from the previous theoretical analysis of the dynamical system presented in this paper. When the propagation is parallel, that is $\theta=0$, the electric field can be eliminated by making a transformation to the wave-frame, resulting in the particle's dynamics being resolved entirely. The particle can be phase-trapped but never physically trapped. When the propagation angle increases, the obliquity becomes manifest through the appearance of a Lorentz-invariant parallel electric field. This electric field physically traps orbits and can result in the creation of a beam parallel to the background magnetic field as well as anisotropies in temperature. Indeed, the oblique propagation can provide an explanation for kinetic distortions of distribution functions for relativistic energies, in a similar manner that it does for the classical case. As the propagation angle increases, the stable fixed point ($\theta<\theta_c$), responsible for trapped orbits and kinetic distortions of distribution functions, goes as $P=\tan(\theta)$ and therefore shifts trapped cells to higher parallel velocities. If the stable fixed point is too distant from the tail of the distribution, no particles will be trapped. This transition from physically trapped to untrapped orbits is singularized by the treatment of the purely perpendicular case. For $\theta=-\pi/2$, as well as for particles that do not belong to the basin of the stable fixed point for $\theta\neq -\pi/2$, the dynamics of the orbit can be simplified to a back and forth slushing on the wave with no net gain in energy on average. \\
On the other hand, if the propagation angle reaches the critical value $\theta=\theta_c$, at which the stability of the fixed point is destroyed by the Hopf bifurcation, the particle belonging to the basin of attraction will be accelerated uniformly to relativistic energies. For $\theta>\theta_c$, a particle initially belonging  to the basin of attraction now becomes chaotic and physical trapping is lost. \\
In-between the regions of phase-space composed of physically trapped and surfing orbits, resides one further source of particle energization. The acceleration in this case originates in the cyclotron-resonance and results in stochastic trajectories.  The inclusion of obliquity as well as the preservation of nonlinearities and relativistic effects, reveal that for a given propagation angle, there is a window in parameter space for which a particle can be accelerated in a diffusive manner primarily along the perpendicular direction. This stochastic process is similar to that of the overlapping resonances for the classical case for obliquely propagating electrostatic waves \cite{Smith78}. The results described in the section above consist indeed of an overlapping of resonances, but does operate for wave-amplitudes about two orders of magnitude lower than those previously assumed. The explanation for this discrepancy with the classical regime is that as the particle gains energy, the dynamical gyrofrequency $\Omega_0=eB_0/m\gamma c$ decreases sufficiently to allow the particle to wander from one resonance to another.\\
The acceleration mechanisms described above both have the important and interesting particularity to operate on short kinetic time scales. The difference is that one operates stochastically and energizes particles primarily along the perpendicular direction, while the second results in a locking in pitch angle and gyrophase, and accelerates particles coherently and primarily along the parallel direction. \\ 
\subsection{Applications to planetary radiation belts.}
The most recent waveforms measured in the radiation belts have revealed an unexpected discovery. Large-amplitudes, monochromatic, obliquely propagating, and bursty waveforms were  not only repeatedly measured in the radiation belts \cite{Catell08, Kellog10, Kersten11, Wilson11}, but appeared correlated with electron energization\cite{Wilson11} as well as relativistic microbursts events \cite{Kersten11}. The correlation between chorus waves and electron energization in the radiation belts is not recent, but it is suspected that if such waveforms were more commonly present in the radiation belts that they could be the dominant trigger responsible for the energization of electrons on short time-scales. A study by Yoon\cite{Yoon11} has shown that if one solves the plasma equations self-consistently, that such waveforms were indeed capable to accelerate electrons on kinetic time scales consistently with the observations. Even though our study lacks the levels of self-consistency provided by the numerical method developed by Yoon\cite{Yoon11}, we arrive to similar conclusions if we choose parameters consistent with the radiation belts measured waveforms. If we integrate the dynamical system for a few wave-periods, and with low-frequency $\delta_2=0.1$, large amplitude $\delta_1\sim0.06$ and for propagation angles obeying the Hopf bifurcations, we find that keV electrons commonly found in the radiation belts could be accelerated on the order of the milliseconds to MeV energies.\\
However, despite encouraging results, we would like to leave a few words of caution. We can not rule that such a mechanism is at play in the radiation belts and the reasons are as follows. 1) There is no clear understanding of the origin of the observed large-amplitude oblique waveforms in the radiation belts. Before we can pinpoint their origin, it is impossible to attempt any self-consistent approach to the current problem. 2) The observations of these waveforms are plagued by uncertainties large enough to seriously undermine any attempt to determine precisely one or multiple acceleration processes. In the very case of the surfatron at Landau resonance, one would need good resolution for the electric and magnetic field components of the waves to obtain propagation angles and wave vectors.  3) Finally, the wave forms are observed with an electrostatic component and the analysis above needs to be conducted with the addition of this compressive electric component. Even though it can be shown that the addition of the electrostatic field with the same phase as the electromagnetic components of the fields would result in the same condition for the surfatron process, a difference in phase would shift the Hopf bifurcation and have non-trivial effects that needs to be scrutinized.\\
In such context, we can not claim that such a mechanism is at play in the radiation belts, but we do suggest that since electrons with keV energies can be accelerated to MeV energies on kinetic timescales, that such mechanism could possibly arise in the radiation belts and other space and cosmic plasmas who are suspected to be permeated by equivalent large-amplitude waveforms \footnote[16]{The stochastic process for the cyclotron resonance case is unlikely to be relevant to the Earth's radiation belts since the amplitudes and phase-speed of the waves are much larger than those observed, but could be of interest in the context of cosmic rays where growing evidence of a structured spectrum suggests multiple acceleration mechanisms.}. 
 
%One short chapter to trace back the work on the radiation belt. One short for the problem of the CR. 
%Recent observations of wave in the radiation belt has shown that waves of such properties not only exist, but appears to be far more common than expected. Indeed....have demonstrated that ..... waves of ..... are.....\\
%Also, even though cosmic plasmas do not benefit from equivalent high resolution measurements, the growing evidence that charged particles can be accelerated on short time-scales by wave-particle processes in plasmas such as the radiation belt, sustains further ground in the assumption that kinetic processes could be holding primordial roles in the understanding of macroscopic and large scale structures deduced from cosmic plasmas.  

\section{Conclusion.}
We have developed a dynamical system to model the interaction of an ion with an obliquely propagating electromagnetic wave in the relativistic limit. We have given a particular focus on the effect of the obliquity on the particle dynamics. It was demonstrated that physical trapping of Landau resonant particles could be identified by the fixed points analysis. Perhaps the main conclusion of our study is that the wave-particle interaction of a single wave demonstrates a rich diversity of mechanisms (acceleration, surfing, stochasticity, trapping) for which the propagation angle is an important and critical parameter. Indeed, the most telling observation, is that the physics at one propagation angle $\theta$ can be significantly altered  for an angle $\theta\pm\epsilon$. \\
Even though the prime difference between oblique propagations with parallel and perpendicular propagations, resides in the inclusion of a region of phase-space for which particles are physically trapped, we have shown that the relativistic treatment also translates in  coupled values in $(\theta, v_\Phi)$ for which particles are accelerated to relativistic energies on kinetic time-scales $\Omega_0\tau\leq1$. Such a mechanism, even though requiring specific wave-properties, can be efficient since it operates on short-time scales, and the volume encompassed by the attractor is large enough to affect a non-negligible portion of a distribution function.\\
Furthermore, it was shown that relativistic effects enhance the cyclotron-resonant stochastic acceleration. As a result of the overlapping in resonances, particles can wander through multiple resonances resulting in a stochastic increase in energy. This relativistic effect is of interest, since it provides acceleration for wave-amplitudes lower than those required for classical regimes of overlapping cyclotron-resonance. Such mechanism could pertain and be more spread than initially assumed in weakly collisional plasmas where particles can be confined for long-time scales.\\
It should finally be pointed out that the model we used is not self-consistent, and will therefore require corrections in order to take into account the complexity of space and astrophysical plasmas. Among these necessary corrections, the departure from a monochromatic spectrum to one composed of a bandwidth, appears today as the most fundamental of them all. Even though some of the large-amplitude waves recently measured in the radiation belt show a significant degree of monochromaticity, the cosmic and space plasmas are mostly turbulent, and the inclusion of additional waves to confirm or infirm the nature of the processes responsible for the acceleration of particles is an inevitable step. However, the dynamical system approach offers numerous advantages and the endeavors for greater self-consistency can be achieved accordingly. Indeed, the dynamical system for the general case, once families of solutions have been found, can be used as a background nonlinear solution to the wave-particle interaction, upon which corrections, such as addition of waves, changing polarization, dispersion effects, inhomogeneous background magnetic field, etc., can all be treated as perturbations to the family of solutions of the "nonlinear homogeneous" system. Such method could be investigated theoretically and numerically, in the similar methodological fashion and with comparable tools that Hamiltonian systems were constructed to investigate the impact of nonlinear perturbations.

\acknowledgments
We thank K. Meziane and L.B. Wilson III for helpful discussions. This work was supported by the Natural Sciences and Engineering Research Council of Canada (NSERC). One of the authors, A. M. H., wishes to acknowledge CSA (Canadian Space Agency) support. Computational facilities are provided by ACEnet, the regional high performance computing consortium for universities in Atlantic Canada.  \\

\appendix

\section{Fixed Points Analysis.}
Fixed points of an $n$-dimensional dynamical system denote stationary solutions for all $n$ variables. Fixed points are defined for values of the variables for which the dynamical system equations equal zero. In this particular case, fixed points represent orbits of physically trapped particles. In order to find the fixed points we proceed as follows. It is clear at first, that in order to have $\dot\gamma=0$, one needs to have $F(\alpha, \Phi, Z)=0$. Keeping this in mind, we first transform the dynamical system as represented by equation sets (\ref{eq:DS}) into polynomial form by using the following change of variables :
\begin{equation}
x=e^{i\alpha}; \hspace{4mm}y=e^{i\Phi};\hspace{4mm}z=e^{iZ};
\end{equation}
and writing the different trigonometric functions in terms of $(x,y,z)$. For the time evolution of $P$ the equation in terms of the $(x,y,z)$ variables result in 
\begin{equation}
(e^{i\alpha}-e^{-i\alpha})(e^{i\Phi}+e^{-i\Phi})\sin(\theta)=0
\end{equation}
\begin{equation}
(x-\frac{1}{x}) (y+\frac{1}{y})=0
\end{equation}
\begin{equation}
\label{eq:FP1}
(x^2-1)(y^2+1)=0
\end{equation} 
since ${x,y,z} \ne 0$. We apply the same procedure for the remaining three equations of motion and we find that for $\dot\alpha=0$ the polynomial equation gives : 
\begin{equation}
\label{eq:FP3}
\left\{ 
\begin{array}{l l} 
\sin(\theta)(x^2+1)(y^2+1)/\delta_2\\
-\delta_1\delta_3 Px[\cos(\theta)(z^2+1)(y^2+1)+(z^2-1)(y^2-1)]=0
\end{array} 
\right.
\end{equation}
the polynomial equation for $\dot\Phi$ is : 
\begin{equation}
\left\{ 
\begin{array}{l l} 
-4P(x^2-1)yz\\

+\sin(\theta)[2\delta_1 P(x^2-1)(z^2+1)y-4x(y^2-1)z/\delta_2\delta_3]\\
+\delta_1 P(x^2+1)[(y^2+1)(z^2-1)+\cos(\theta)(y^2-1)(z^2+1)]=0
\end{array} 
\right.
\end{equation}
and the polynomial equation for  $\dot{Z}$ resumes as : 
\begin{equation}
\label{eq:FP2}
2\cos(\theta)y(x^2+1)-\sin(\theta)(x^2-1)(y^2-1)=0
\end{equation}
The equation (\ref{eq:FP1}) has the following solutions : 
\begin{equation}
x=\pm1 ; y=\pm i
\end{equation}
We look at each solution starting with the case $x=\pm 1$. Replacing $x$ in equation (\ref{eq:FP2}) cancels the second term and results in the following constraint : 
\begin{equation}
4\cos(\theta)y=0
\end{equation}
Since the solutions of this equation are 
\begin{equation}
\cos(\theta)=0; \hspace{5mm} y=0.
\end{equation}
neither of them is acceptable. We are interested in the case of oblique propagation with $\theta\ne \frac{\pi}{2}$ and  by definition $y\ne 0$. Hence the case $x=\pm1$ does not result in a fixed point for the oblique propagation.

We now look at the second solution which satisfies equation (\ref{eq:FP1}), $y=\pm i$. We replace $y$ in equation (\ref{eq:FP2}) and find
\begin{equation}
\pm i \cos(\theta)(x^2+1)+\sin(\theta)(x^2-1)=0
\end{equation}
\begin{equation}
x^2(\pm i \cos(\theta)+\sin(\theta))=-(\pm i \cos(\theta)-\sin(\theta))
\end{equation}
\begin{equation}
x^2(\pm(e^{i\theta}+e^{-i\theta})-(e^{i\theta}-e^{-i\theta}))=-(\pm(e^{i\theta}+e^{-i\theta})+(e^{i\theta}-e^{-i\theta}))
\end{equation}
which results in 
\begin{equation}
\label{eq:FP4}
x^2=-e^{\pm2i\theta}; \hspace{5mm} x=\pm i e^{\pm i\theta}
\end{equation}
To find the value of $z$ we replace the value of $y$ in the equation (\ref{eq:FP3}) which results in the cancelation of  the first two terms and gives the following result : 
\begin{equation}
-2\delta_1 Px (z^2-1)=0
\end{equation}
Since by definition $x\ne0$ and we are not interested with the trivial case $P=0$ the solution for this equation is :
\begin{equation}
z=\pm1
\end{equation}
The last step consists in finding the value of P for the fixed point, which can be done by replacing $y=\pm i$ and $z=\pm 1$ in equation (\ref{eq:FP2}).
\begin{equation}
\left\{ 
\begin{array}{l l} 
-4P(\pm i)(\pm 1)(x^2-1)+4 P\sin(\theta)\delta_1(\pm i)(x^2-1) \\
+8\sin(\theta)(\pm 1)x/\delta_2\delta_3-4\delta_1P\cos(\theta)(x^2+1)=0
\end{array} 
\right.
\end{equation}
which can be written as : 
\begin{equation}
-P=\pm 2i\frac{x}{x^2-1}\frac{\sin(\theta)}{\delta_2\delta_3}
\end{equation}
Replacing $x$ with the use of equation (\ref{eq:FP4}), we find : 
\begin{equation}
\delta_2\delta_3P= -\tan(\theta)
\end{equation}
Since $P$ is always positive the solution requires $\frac{-\pi}{2}<\theta<0$. Transforming back $x,y,z$ to the dynamical system variables $\alpha, \Phi, Z$ we find that the fixed points are given by: 
\begin{eqnarray}
\label{eq:FP5}
P=-\gamma \tan(\theta); \hspace{10mm} \alpha =\pm\theta\pm\frac{\pi}{2}; \nonumber \\
\Phi=\pm \frac{\pi}{2}; \hspace{15mm} Z=0,\pi.
\end{eqnarray}
Using the values in equations (\ref{eq:FP5}) for $F(\alpha, \Phi, Z)$ sets it equal to zero. Hence, equations (\ref{eq:FP5}) are the fixed point equations for the dynamical system (\ref{eq:DS}). We can also express the fixed points in terms of $(p_x, p_y, p_z, z', \gamma)$ as demonstrated in table 1. 
\begin{table}[!h]
\caption{Fixed points in the $(p_x, p_y, p_z, Z, \gamma)$ representation }
\begin{center}
\begin{tabular}{| l | c | c | c | c |}
\hline
$p_{x0}$&$p_{y0}$&$p_{z0}$&$Z_0$&$\gamma_0$\\
\hline\hline
$0$&$-p_\Phi\tan(\theta)$&$p_\Phi$&$0$&\Large{$\frac{1}{\sqrt{1-\frac{v_\Phi^2}{c^2}(1+\tan^2(\theta))}}$} \\
\hline
$0$&$-p_\Phi\tan(\theta)$&$p_\Phi$&$\pi$&\Large{$\frac{1}{\sqrt{1-\frac{v_\Phi^2}{c^2}(1+\tan^2(\theta))}}$} \\\hline
$0$&$+p_\Phi\tan(\theta)$&$p_\Phi$&$0$&\Large{$\frac{1}{\sqrt{1-\frac{v_\Phi^2}{c^2}(1+\tan^2(\theta))}}$} \\
\hline
$0$&$+p_\Phi\tan(\theta)$&$p_\Phi$&$\pi$&\Large{$\frac{1}{\sqrt{1-\frac{v_\Phi^2}{c^2}(1+\tan^2(\theta))}}$} \\
\hline
\end{tabular}
\end{center}
\end{table}

\section{Stability Analysis} 
The next fundamental step in dynamical system theory is to investigate the equilibrium of the fixed points. In order to do so, we apply a basic Lyapunov linear analysis that can be found in any textbook on dynamical systems. The method is summarized as follows. For a dynamical system $\mathbf{\dot{x}}=F(\mathbf{x})$ possessing a fixed point $\mathbf{x_0}$ for which $F(\mathbf{x_0})=0$, one can make a Taylor expansion around the fixed point and keep the first order terms. That is, writing the dynamical for the perturbation $\mathbf{\delta x}$ and the Jacobian $\mathbf{J}=\frac{\partial F}{\partial x_i}\bigg{|}_{\mathbf{x}_0}$ as  $\frac{d}{dt}\mathbf{{\delta x}}=\mathbf{J\delta x}$. We are left with the task of solving an eigenvalue problem since we can write the solution to the linearized equation as  $\mathbf{\delta x}=\mathbf{\xi}_ie^{\lambda_i t}$ for the eigenvalues $\lambda_i$ and eigenvectors $\xi_i$, assuming the eigenvalues are not degenerate. If one eigenvalue $\lambda_i > 0$ the system is linearly unstable at $\mathbf{x_0}$, if not the system is linearly stable at $\mathbf{x_0}$.
From the expression for $\gamma_0$ in the previous appendix, it is clear that a fixed point does not exist for all parameter values of $\theta$ and $v_\Phi$. Since $\gamma \geq 1$, the argument in the denominator square root must  obey the condition $n^2\geq 1+\tan^2(\theta)$. 
Hence we write the Jacobian for the dynamical system as follows : 
\[
 \left( \begin{array}{cccc}
0 & \frac{\cos(\theta)}{\delta_2\gamma_0} &\frac{\mp\delta_1+\sin(\theta)}{\delta_2\gamma_0} &0 \\
- \frac{\cos(\theta)}{\delta_2\gamma_0}  & 0 & 0&0 \\
(\frac{-\sin(\theta)}{\delta_1}\pm\frac{n^2-1}{n^2})\frac{\delta_1}{\delta_2\gamma_0} & 0 & 0&\mp\frac{\delta_1\tan(\theta)}{\delta_2}\frac{n^2-1}{n^2}\\
0&0&\frac{1}{\gamma_0}&0 \end{array} \right),\]
where the $\pm$ symbols denote the two values of the fixed points in $Z=kz'$. Solving the eigenvalue problem $(\mathbf{J}-\lambda\mathbf{I})\mathbf{\xi}=0$ we find a bi-quadratic polynomial function in $\lambda$ that can be written as $\chi(\lambda)=\lambda^4+\eta_1\lambda^2+\eta_2=0$, with the constant coefficients $\eta_1$ and $\eta_2$ given by the following expressions : 
\begin{eqnarray}
\eta_1&=&\frac{\delta_1}{\delta_2\gamma_0}\frac{n^2-1}{n^2}\tan(\theta)+\frac{\cos^2(\theta)}{\delta_2^2\gamma_0^2}\nonumber\\
  &-&\frac{\delta_1}{\delta_2^2\gamma_0^2}\bigg{(}-\frac{n^2-1}{n^2}\pm2\sin(\theta)-\frac{\sin^2(\theta)}{\delta_1}\mp\frac{\sin(\theta)}{n^2}\bigg{)}\nonumber\\
 \eta_2&=&\frac{\delta_1}{\delta_2^2\gamma_0^2}\frac{n^2-1}{n^2}\sin(\theta)\cos(\theta)
\end{eqnarray}
A close look at the coefficients of equation set $(7)$ shows that all four eigenvalues will cross the zero real axis when the condition
 \begin{equation}
 n^2-1=\tan^2(\theta)
 \end{equation}
 is respected. That is, for parameter values corresponding to $\gamma_0^{-1}=0$ and resulting in $\lambda^4=0$.

% HERE IS THE NAME OF THE BIBLIOGRAPHY .BIB FILE.
% IT PUTS REFERENCES AUTOMATICALLY INTO APS STANDARD FORMATS.

%\bibliography{owpi3}

\end{document}